# High Temperature Emissivity, Reflectivity, and X-ray absorption of BiFeO$_3$

Néstor E. Massa[*]
Laboratorio Nacional de Investigación y Servicios en Espectroscopía Óptica-Centro CEQUINOR, Universidad Nacional de La Plata, C.C. 962,
1900 La Plata, Argentina,

Leire del Campo,[§] Domingos De Sousa Meneses, and Patrick Echegut,
CNRS-Conditions Extrêmes et Matériaux Haute Température et Irradiation,
1D, Av. de la Recherche Scientifique, F-45071 Orléans, France,

Gilberto F. L. Fabbris
Laboratório Nacional de Luz Sincrotron (LNLS)
and
Instituto de Fisica, Universidade Estadual de Campinas, Campinas CEP 13083-970,SP, Brazil

G. de M. Azevedo
Laboratório Nacional de Luz Sincrotron (LNLS)
and
Instituto de Física, Universidade Federal do Rio Grande do Sul, Av. Bento Gonçalves, 9500, Caixa Postal 15051, 91501-970 Porto Alegre, RS, Brazil

María Jesús Martínez-Lope, and José Antonio Alonso
Instituto de Ciencia de Materiales de Madrid, CSIC,
Cantoblanco, E-28049 Madrid, Spain.

---

- e-mail : neemmassa@gmail.com

# ABSTRACT

We report on the lattice evolution of $BiFeO_3$ as function of temperature using far infrared emissivity, reflectivity, and X-ray absorption local structure.

A power law fit to the lowest frequency soft phonon in the magnetic ordered phase yields an exponent β=0.25 as for a tricritical point. At about 200 K below $T_N$~640 K it ceases softening as consequence of $BiFeO_3$ metastability. We identified this temperature as corresponding to a crossover transition to an order-disorder regime. Above ~700 K strong band overlapping, merging, and smearing of modes are consequence of thermal fluctuations and chemical disorder. Vibrational modes show band splits in the ferroelectric phase as emerging from triple degenerated species as from a paraelectric cubic phase above $T_C$~1090 K. Temperature dependent X-ray absorption near edge structure (XANES) at the Fe K-edge shows that lower temperature $Fe^{3+}$ turns into $Fe^{2+}$. While this matches the FeO wüstite XANES profile, the Bi $L_{III}$-edge downshift suggests a high temperature very complex bond configuration at the distorted A perovskite site. Overall, our local structural measurements reveal high temperature defect-induced irreversible lattice changes, below, and above the ferroelectric transition, in an environment lacking of long-range coherence. We did not find an insulator to metal transition prior to melting.

# I-INTRODUCTION

Magnetoelectrics, the field for compounds known under the generic name of multiferroics has as main theme the searching for materials developing ferroelectric polarization in a magnetic ordered phase.[1] More than 100 years ago P. Curie pointed out correlations in low symmetry crystals magnetic and electric properties "c´est la dissymmétrie qui crée le phènoméne".[2] Later reformulated by Landau and Lifshitz, as "the magnetoelectric effect is odd with respect to time reversal and vanishes in materials without magnetic structure"[3] the field received a new boost with the rediscovery of materials that show spontaneous ferroelectricity in addition to magnetic ordering.[4] Currently, they are understood as having two order parameters, spontaneous polarization (antiferroelectric, ferroelectric, ferrilectric), and spontaneous magnetization (antiferromagnetism, ferromagnetism, ferrimagnetism) triggering, by magnetoelectric coupling, one order by the other. It is then due, anticipating the envisioned potential on applied techniques, a proper and full understanding of the lattice dynamics of potential multiferroics and their precursor phases.

Among compounds called multiferroic, $BiFeO_3$ is distinct in that a ferroelectric and an antiferromagnetic phase transition take place above room temperature.[5] $BiFeO_3$ may be identified with the simple ferroelectric perovskite structure, it belongs to the group compounds having at the A site two valence electrons participating in chemical bonds through (sp)-hybridized states such as $sp^2$ or $sp^3$. These valence electrons in s-orbitals are the "lone pair" electrons. The lone pair is unstable, mixing the $(ns)^2$ ground state and a low lying $(ns)^1(np)^1$ excited state, and thus, leading the ions to break the lattice inversion symmetry yielding ferroelectricity.[6] Then, the two 6s shell electrons of $Bi^{3+}$ and the magnetic $Fe^{3+}$ ions, at the A and B perovskite sites respectively, yield a compound within the framework of "proper" ferroelectrics allowing magnetism and ferroelectricity.

Earlier work on bismuth ferrite $BiFeO_3$ remounts to a report by Smolenski and Chupis.[7,3] Palai et al[8] using several structural and optical techniques proposed the rhombohedral R*3c* -α- phase for temperatures below $T_C$, an intermediate paraelectric -β- phase between 1130 to 1225 K, and a cubic -γ- phase above at T ~1225-1233 K, coincident with an hypothetically high temperature insulator-metal phase transition. $BiFeO_3$ in the rhombohedrally distorted phase has ferroelectric, antiferromagnetic, and ferroelastic ordering above 300 K.[9] It orders antiferromagnetically below $T_N$ ~640 K in a spiral spin incommensurate structure [10,11,12,13] within the ferroelectric phase[14,15] found below $T_C$~1090 K associated with large atomic displacements. Haumont et al[15] detected anomalies in the lattice parameters at $T_N$ and proposed a monoclinic $P2_1/m$ space group for the high temperature paraelectric phase. More recently, it was recognized by neutron powder diffraction patterns that the -β- $BiFeO_3$ phase is orthorhombic, $GdFeO_3$–P*bnm*($D_{2h}^{16}$) isostructural, coexisting with the lower temperature $R3c$ α-phase in the 1120 – 1130 K range thus explaining the order-disorder character found at $T_C$ ~1090 K.[16] These measurements suggest the higher temperature paraelectric β phase persisting up to complete decomposition of $BiFeO_3$ into $Bi_2Fe_4O_9$ in coincidence with R. Przenioslo et al that reported significant decomposition into $Bi_2Fe_4O_9$ at 973 K.[17]

First principles studies on the spontaneous polarization of $BiFeO_3$ were first advanced by Neaton et al.[18] Ranvindran et al[19] predicted the structural, electronic, magnetic, and ferroelectric properties of $BiFeO_3$ based upon density functional calculations. They conclude that a large portion of the ferroelectric polarization is provided by Bi displacements relative to the center of the $FeO_6$ octahedra with the Bi lone pair causing symmetry lowering. While large Bi-O and Fe-O hybridization plays a dominant role in the ferroelectric polarization, the overall behavior of $BiFeO_3$ is found to belong to the ferroelectric displacive type. As we will see below this picture holds better below room temperature.

Factor group analysis for BiFeO$_3$ (Z=2) room temperature R$3c$ (C$^6_{3v}$) space group yields the irreducible representation Γ = 5 A$_2$ + 4 A$_1$+ 9 E, thirteen A1 and E optical modes, being Raman and infrared active. In the higher temperature paraelectric phase there are only three infrared active modes, Γ= 3F$_{1u}$ + 1 F$_{2u}$, if the cubic phase Pmcm (Z=1) is adopted.[20] The orthorhombic GdFeO$_3$ distortion (Z=4) more recently proposed by Arnold et al[16] allows Γ=7B$_{1u}$+9B$_{2u}$+9B$_{3u}$. infrared active phonons.[21]

There are a number of Raman scattering measurements in the literature. Among them, Haumont et al[22] pointed a cubic Pm¯3m structure for the high temperature paraelectric phase and departures in the expected soft-phonon behavior at lower temperatures. Fukumura's et al[23] single crystal measurements allowed phonon identification as predicted from group theory and speculate on phonon-magnetic order coupling. Polar phonons and spins excitations coupling up to the Néel temperature, suggesting magnetostriction and piezoelectric effects, were reported by Rovillain et al.[24] Low temperature infrared and Raman active phonons were also calculated from first principles by Hermet et al[25] and by Tütüncü and Srivastava.[26]

Here we discuss our measurements for BiFeO$_3$ combining far infrared reflectivity and emission spectroscopy putting emphasis on the high temperature ferroelectric to paraelectric poorly understood phase transition at $T_C \sim 1090$ K.

Since the resulting far infrared band profiles suggest a complex panorama, to improve our understanding on the lattice dynamics, we also probed the local structure getting information in an additional length scale. Temperature dependent Bi and Fe X-ray absorption near edge structure (XANES) spectra allows inferring ion valence states against the infrared view just below, at, and above the high temperature structural transition at $T_C \sim 1090$ K. We find irreversible lattice changes on warming revealing overall instabilities and lack of lattice long-range coherence at temperatures well below 1000 K. This last may be interpreted as due to inhomogeneities, that departing from $BiFeO_3$ stoichiometry and structure, creates a complex environment reminiscent to ferroelectric relaxors.[27] We conclude that at temperatures higher than ~800K it may be inappropriate to strictly assign a space group to $BiFeO_3$ in spite of detecting the ferroelectric-paraelectric phase transition at $T_C$. We also show that, although we cannot entirely rule out an overdamped Drude component hidden in the lowest frequency part of the spectrum, we do not detect an insulator to metal transition prior to melting.[28]

## II-EXPERIMENTAL DETAILS

Optical as well as X-ray absorption measurements were performed using the same batch of samples. Polished high quality polycrystalline samples in the shape of 5mm diameter BiFeO$_3$ pellets were prepared in polycrystalline form from citrate precursors obtained by a soft chemistry procedure. Stoichiometric amounts of analytical-grade Bi$_2$O$_3$, FeC$_2$O$_4$.2H$_2$O were dissolved in citric acid, adding some droplets of HNO$_3$ to facilitate the dissolution of the bismuth oxide; the citrate solution was slowly evaporated, leading to an organic resin which was dried at 120ºC and then slowly decomposed at temperatures up to 500º C in air. The samples were slowly heated up to 800ºC, and held at this temperature for 2 hrs.

BiFeO$_3$ was obtained as a well crystallized powder, as shown in Fig. 1, where the splitting of the (104) and the (110) reflections at about 32º is perfectly distinguished. The crystal structure was refined from the X-ray diffraction data at ambient temperature in the conventional R*3c* model, with Bi at 6a (0,0,0) sites, Fe at 6a (0,0,z), positions with z~0.223 and O at 18b (x,y,z) sites, with x~0.437, y~ 0.022, z~0.959. The unit cell parameters are a=5.582(1), c= 13.869(4) Å in the hexagonal setting, in good agreement with those published elsewhere (for instance, a= 5.58102(4), c= 13.8757(2) Å in ref [29]).

Transmission-mode X-ray absorption measurements were done at the D04B-XAFS1 beamline in the Brazilian Laboratory for Synchrotron Light (LNLS).[30] Channel-cut Si monochromators (220 and 111 reflections, respectively) were utilized for measurements around bismuth $L_3$ edge (13419 eV) and iron K-edge (7112 eV). Samples were prepared by mixing powder of $BiFeO_3$ with boron nitride and pressing the mixture into a pellet. Different pellets were prepared, with amounts of $BiFeO_3$ and BN calculated so as to yield an edge jump~1 near the Fe K-edge or Bi $L_3$-edge. During the measurements, the samples were placed inside a furnace and the temperature was controlled in the 300 K-1120 K range. Data analysis was performed using the IFEFFIT package.[31]

Temperature dependent medium (MIR), and far (FIR) infrared near normal reflectivity in the 10 $cm^{-1}$-5000 $cm^{-1}$ spectral region were measured in a FT-IR Bruker 113v interferometer with 1 $cm^{-1}$ resolution. A gold mirror was used as 100% reference. Our samples were mounted on a cold finger of a home-made cryostat for measurements between 4 K and 300 K. For high temperatures up to ~600 K we used a heating device adapted to the Bruker 113v vacuum chamber and the near normal reflectivity attachment. In this temperature range, the reflectivity was corrected from the spurious infrared signal introduced by the hot sample thermal radiation. Overlapping, and at higher temperatures, we measured near normal spectral emissivity, with 2 cm$^{-1}$ resolution, in a facility that allows far- and mid- infrared measurements by making use of two interferometers, a Bruker Vertex 80v and a Bruker Vertex 70. Both instruments are simultaneously optically coupled to a rotating table through a dry air box where the sample is positioned on a ceramic holder. The sample was heated from the bottom with a 500 W pulse Coherent $CO_2$ laser allowing heating between 600 K and 1300 K.

All our measurements, except when explicitly stated, were taken on heating runs.

The normal spectral emissivity of the sample, $\mathscr{E}(\omega,T)$, is given by the ratio of its luminescence ($\mathscr{L}_S$) relative to the black body's ($\mathscr{L}_{BB}$) at the same temperature T and conditions, thus,

$$\mathscr{E}(\omega,T) = \frac{\mathscr{L}_S(\omega,T)}{\mathscr{L}_{BB}(\omega,T)} \qquad (1)$$

In practice, the evaluation of this quantity needs the use of a more complex expression (eq. 2) because the measured fluxes are polluted by parasitic radiation due to the fact that part of the spectrometer and detectors are at 300 K. To eliminate this environmental contribution the sample emissivity is retrieved from three measured interferograms, i.e., sample, $I_S$; black body, $I_{BB}$; and, environment, $I_{RT}$; by applying the following relation

$$\mathscr{E}(\omega,T) = \frac{FT(I_S - I_{RT})}{FT(I_{BB} - I_{RT})} \times \frac{\mathscr{P}(T_{BB}) - \mathscr{P}(T_{RT})}{\mathscr{P}(T_S) - \mathscr{P}(T_{RT})} \mathscr{E}_{BB} \qquad (2)$$

where $FT$ stands for Fourier Transform., $\mathscr{P}$ is the Planck's function taken at different temperatures T; i.e., sample, $T_S$; blackbody, $T_{BB}$; and surroundings, $T_{RT}$. $\mathscr{E}_{BB}$ is a correction that corresponds to the normal spectral emissivity of the black body reference (a $LaCrO_3$ Pyrox PY 8 commercial oven) and takes into account its non-ideality. The blackbody temperature, $T_{BB}$, was kept at 1273 K for stability reasons. This is why a supplementary Plank´s correction is used for taking into account the effects of temperature difference between sample and reference.[32,33]

Emissivity allows contact free measurement of the sample temperature using the Christiansen frequency; i.e., the frequency where the refraction index is equal to one, and the extinction coefficient is negligible, just after the highest longitudinal optical phonon frequency. The temperature is calculated using eq. (2) with the emissivity $\mathscr{E}(\omega,T)$ at that frequency set equal to one.

After acquiring the optical data we placed all our spectra in a more familiar near normal reflectivity framework by noting that

$$\mathscr{R} = 1 - \mathscr{E} \quad (3)$$

where $\mathscr{R}$ is the sample reflectivity. Then, we computed phonon frequencies using a standard multioscillator dielectric simulation.[34] The dielectric function, $\varepsilon(\omega)$, is given by

$$\varepsilon(\omega) = \varepsilon_1(\omega) + i\varepsilon_2(\omega) = \varepsilon_\infty \prod_j \frac{(\omega_{jLO}^2 - \omega^2 + i\gamma_{jLO}\omega)}{(\omega_{jTO}^2 - \omega^2 + i\gamma_{jTO}\omega)} \quad (4)$$

from where we calculate the normal reflectivity using the Fresnel formulae optimized against the experimental data. $\varepsilon_\infty$ is the high frequency dielectric constant taking into account electronic contributions; $\omega_{jTO}$ and $\omega_{jLO}$, are the transverse and longitudinal optical mode frequencies and $\gamma_{jTO}$ and $\gamma_{jLO}$ their dampings, respectively.[35] In addition, we calculated the oscillator strength $S_j$ for the $j^{th}$ oscillator as in

$$S_j = \omega_{jTO}^{-2} \frac{\prod_k (\omega_{kLO}^2 - \omega_{jTO}^2)}{\prod_{k \neq j}(\omega_{kTO}^2 - \omega_{jTO}^2)} \quad (5)$$

Table I shows that at 4 K the number oscillators necessary in the fit is higher than those room temperature space group allowed. Above 300 K, it gradually decreases until 523 K, and higher temperatures, where nine are enough to obtain excellent spectral fits.

## III-RESULTS AND DISCUSSION

Fig. 2(a, b, c, d, e ) shows the far infrared reflectivity spectra of $BiFeO_3$ in the entire temperature range from 4 K to melting. The lower temperature spectra are shown in Fig. 1a where the thirteen infrared active phonon bands predicted for the ferroelectric phase at 300 K by group analysis are individualized. The spectra agree with those already reported in the recent literature by Kamba at al[20] and Lobo et al[36]. We also noted that at 4 K we could improve the reflectivity fit by including three extra shoulders (Table I) that might point to a lower temperature structural distortion.

Fig. 2b shares spectra taken by near normal reflectivity and by emission at higher temperatures up to melting. As the temperature increases we observe lattice softening and phonon broadening effectively reducing the number of oscillators needed for fitting (our fits satisfy $\chi^2 \sim 1.\ 10^{-5}$) suggesting a high symmetry far infrared view of the paraelectric phase .

Fig. 2 (c,d,e) and Table I are fitting examples of those spectra. As it is shown in Fig. (2d), we find an excellent absolute correspondence between the spectra taken under near normal reflectivity and those by emission at higher temperatures. Fig. (2e) shows spectra at 1120 K and 1190 K in the paraelectric phase. Only three oscillators (Table I) are needed to achieve an excellent fit in agreement with the number of bands predicted using the cubic Pm¯3m space group.[26] It is worth noticing that in the spectrum at 1190 K, and in spite that at these temperatures decomposition has a significant role, we can still distinguish a slight band hardening and a reduction in the absolute reflectivity. It suggests detecting a paraelectric soft phonon frequency shifting toward higher frequencies in the hypothetical cubic phase. When the temperature is further increased up to 1273 K our samples partially melt, the pellets turning into drops.

Evidence for phonon overlapping and broadening starts above 600 K where the thermal induced anharmonicities and sample inhomogeneities materialize. They blur mode split at ~80 cm$^{-1}$ while the phonon bands centered at 68 cm$^{-1}$ and 130 cm$^{-1}$ soften toward T $\rightarrow$ T$_C$. These bands behave as non-degenerate A1 and double-degenerated E1 modes (fig. 3c) originating in a paraelectric cubic three-fold degenerated F$_{1u}$ specie. In fact, between 1090 K and 640 K, the split into two band pairs may be interpreted corresponding to each of the three $F_{1u}$ three-fold degenerated cubic bands. Thus, the dynamics yields an effective symmetry at molecular level allowing to be simulated by nine wide and overdamped oscillators (Table I) (shorter phonon lifetimes) in response to the occurrence of a number of inhomogeneity islands. The corresponding phonon line-widths are reminiscent to those found in relaxor ferroelectrics.[27] This trend continues up to melting.

The temperature dependence of our spectra at high temperatures suggests phonon behavior as in $PbTiO_3$ (ref. 37-39) with a cubic-order-disorder-displacive phase sequence. In our case the products of the $BiFeO_3$ metastability[40] lead the order-disorder part of the lattice dynamics at high temperatures.

We then conclude that the antiferromagnetic transition temperature at $T_N \simeq 640$ K, may be considered as a crossover temperature above which a quasi-displacive picture driven by phonon softening is no longer valid. This is the temperature where a continuous volume expansion takes place,[40] and where an atomic displacement is associated with a local minimum for the rhombohedral angle.[17]

Nonetheless, the characteristic marker for lattice instability in the ferroelectric phase is the lowest frequency transverse optical mode. As it was also reported in refs. 20 and 36, softening starts even at the lowest temperatures. It has a distinctive half width band increase at room temperature suggesting an intrinsic unstable lattice (Fig. 3b). As the temperature is risen, and in spite that at about 400K-500K ceases softening, it is possible to trace its temperature dependent behavior up to Tc because the dielectric simulation (eq. 4) has independent oscillators as built-in hypothesis. Table I shows the minimum number of fitting parameters satisfying our fitting criterion at each temperature in which the spectra suggest a differentiated effective lattice environment.

Fig. 3d shows that soft phonon peak position in the low temperature regime deviates from a power law at about 100 K below $T_N \simeq 640$ K, the temperature where the spin cycloid disappears.[41] Below we present arguments suggesting that the destruction of the magnetic order is lattice driven.

The power law fit using soft phonon optical conductivity maxima,

$$\omega_{soft} = A \cdot (To - T)^{\beta} \quad , \qquad (6)$$

being $\mathcal{A}$ a constant, and To is an effective ferroelectric transition temperature, yields an exponent, β=0.25. (To ± 50 degrees, with an exponent band error between +0.01 and −0.03). That value, and the low temperature behavior shown as a solid line in fig. 3d, is of critical regimes close to temperatures at which a structural phase transition takes place. It suggests a delicate balance between a discontinuous first order and a continuous second order phase transition. I.e., where the fourth power coefficient, in the free energy expanded in terms of the order parameter, is zero. This is the line where the first order transition meets the line of the second order transition.[42] That β=0.25 holds in BiFeO$_3$ in a large temperature range below T$_N$~640 K, where the lattice has R*3c* (C$^6_{3v}$) symmetry, is indicative that BiFeO$_3$ is strongly affected by structural fluctuations. In other words, the ferroelectric mode driven phase transition that might be deduced from the power law is frustrated by the metastability of BiFeO$_3$ at high temperatures. Here, we find a distorted environment with competing phases in the form of clusters. In particular, it is instructive to look at figure 3 of Carvalho and Tavares (ref. 43) where it is shown that at fixed high temperatures, if one waits long enough, all BiFeO$_3$ will convert into secondary phases associated with the chemical reaction

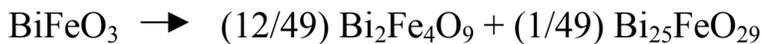

BiFeO$_3$ → (12/49) Bi$_2$Fe$_4$O$_9$ + (1/49) Bi$_{25}$FeO$_{29}$

It is also worth stressing that our conclusions are partially a consequence of far infrared spectral wavelengths at which we are measuring. From ~15 μm to ~1000 μm, homogeneity results from cluster spectral averaging and thus, almost certainly, decomposition products and defects go undetected. And by the same token, the assignment of a minority phase for the decomposition product Bi$_2$Fe$_4$O$_9$ by Arnold et al[16], and in the elaborated fit by Haumont et al[15] for the high temperature paraelectric phase, is to be understood in terms of lattice long-range coherence embedded in the Rietveld analyses of their neutron and X-ray diffraction patterns.

It is then proper to seek other techniques, as X-ray absorption local structure, yielding a perspective in another length scale, as a way to elucidate the characteristics of the order-disorder phenomena in $BiFeO_3$.

The temperature evolution of XANES spectra around the bismuth edge is shown in Fig. 4. Clearly, the amplitude of the fine structure oscillations progressively decreases as the temperature is increased, eventually completely disappearing above 1074 K. At 1123 K there is no visible fine structure oscillation, consistent with a high degree of disorder in the local atomic environment around bismuth. We also note the edge position shifts to lower energies. This reduction may be interpreted as if the valence for Bismuth departs for the nominal $Bi^{3+}$ expected if the only lattice components were cluster inhomogeneities made of $BiFeO_3$, $Bi_2Fe_4O_9$, or $Bi_{25}FeO_{39}$. Rather, the down energy shift detected at the Bi edge and the change in the XANES profiles as temperature increases is reminiscent to what it is known for X-ray absorption of polymorph $Bi_2O_3$ and its metastable phases.[44]

$Bi_2O_3$ has complex structures with Bi ions occupying at least two types of irregular coordination polyhedra where 6d states and 6s states are accessible.

In our case Bi 6s is filled and thus there are no 2p3/2⟶6s transitions. The rest of the observed features are known to be Bi 2p3/2⟶6d transitions.[44] Increasing bond distances result in XANES peak positions shifting to lower energies in a $Bi_2O_3$ environment comparable to the distorted A perovskite site expected at high temperatures for $BiFeO_3$. Then, the presence of irregular polyhedra around Bi above 700 K correlates to the smoothing on heating of ill defined peaks that originate in *p-d* transitions at ~13400 eV in the XANES spectra (Fig. 4). The energy downshift is the net effect of having a distorted cage for Bi. The A perovskite lattice site turns across the paramagnetic phase into complex polyhedra, with many slightly different lattice distances. It is also possible, on the other hand, that the downshift might have contributions of a formal valence state for Bi as in $(Bi_2O_2)^{2+d}$, i.e. $Bi^{\sim+2.75}$, in a local distorted ferroelectric aurivillius-like environment.[45,46] This is in line with irreversible changes in the Bi-O distance distribution and coincident with the reported lattice anomalies above $T_N$ ~ 640 K. [16,17,40]

We have also performed measurements at the Fe K-Edge on warming and cooling. As it is shown in Fig. 5, XANES edge position shifts to lower energies as the temperature increases with the oxidation state of iron reduced to $Fe^{2+}$ from $Fe^{3+}$. Above 1074 K peak position agrees with standard FeO wüstite in a different coordination environment (Fig. 6).There are two possible nearest-neighbor charge transfers in the FeO lattice. This occurs between edge-sharing or corner-sharing $FeO_6$ octahedra with II/III valence interchange between nearest-neighbor Fe atoms and with the Fe(III) constituting a "hole" electronic defect.[47] Heating $BiFeO_3$ seems to enhance $Fe^{3+}/Fe^{2+}$ valence fluctuations of the Fe ion in a rarified environment. At room temperature, this is interpreted in terms of oxygen vacancies that by charge compensation reduce the electrical resistivity and thus the effective electronic gap.[48,49]

And as it was already stated for the Bi edge, when the temperature is lowered to ambient, XANES profiles are not recovered pointing again to irreversible processes likely also involving oxygen loss. These would induce phase separation into secondary phases at the expense of the pure perovskite structure.

The local symmetry breaking, that might had been enhanced by hidden grain size dependence, contrasts with our measurements in the infrared where one can follow an overall perovskite-like phonon activity up to melting. These differences are close to findings in many perovskite related compounds where the average cubic $Pm\bar{3}m$ is detected although X-ray absorption shows tetragonal or orthorhombic distortions on a nanometric scale.[50] However, for explaining the differences found in $BiFeO_3$ using infrared and X-ray absorption, we believe that it seems appropriate to bring also into the analysis the consequences of its high temperature thermodynamics.

We note that recent in- and ex-situ X-ray diffraction patterns confirmed that $BiFeO_3$ is metastable at high temperatures with respect to $Bi_2Fe_4O_9$ and $Bi_{25}FeO_{39}$ making the effective data acquisition time a non-negligible factor. Infrared and Raman spectra are likely to show only small departures from the perovskite-like expected behavior. The reason is that, beyond the number of temperatures in which the spectra were recorded, each run would take about, say, 15 minutes, and thus, even that it is an accumulative measuring process; they do not reach a critical time expanse that would allow secondary phases in unacceptable amounts. The bands measured in the infrared, at temperatures between 600 K and 1200 K must be interpreted as net results since frequencies of minority phases are by their very nature close to those of $BiFeO_3$. That is, minority phase vibrational modes are embedded in the overall infrared spectra. Samples in our X-ray absorption measurements were in an oven where each temperature run takes two and half hours. As it was shown above (figs. 4 and 5), the absorption spectrum has the expected perovskite profile up to around 700 K where spectra retain the perovskite characteristics, and then, at 900 K shows for Bi edge a clear anomaly indicating a strong distortion for A site.

The same conclusion may be reached for the Fe inside distorted oxygen octahedra.

This should be considered in the subsequent understanding of irreversible features found on heating [43,51], thus, helping to reach common grounds of apparent different results.[8,15,16,22]

Finally, as it is shown in Fig. 2b (dotted lines), our spectra do not back the idea of a possible insulator-metal phase transition at atmospheric pressure as proposed by Scott et al.[28] In far infrared, the plasma of a hypothetical metallic phase would be easily detected as a Drude Lorentzian, centered at zero frequency. In metal oxides an increment in the number of freer carriers is also accompanied by strong electron-phonon interactions, carrier hopping, and polarons. This is generally detected as an inflection at about longitudinal optical phonon frequencies and/or as an overdamped oscillator in the mid-infrared.[52] Our reflectivity spectra in the 1090 K to 1279 K range does not show neither a reflectivity jump at low frequencies nor any electron-phonon related contribution expected for the increment on the number of carriers in a –γ- metallic phase.[8] And although we can not entirely rule out the possibility of an overdamped Drude contribution hidden in the reflectivity at very low frequencies, our spectra is composed of unstructured smoothed bands behaving as those for a regular insulator toward melting beyond the peritectic decomposition temperature at 1207 K.[53] This conclusion is also in agreement with neutron scattering data up to 1173 K suggesting the absence of the –γ-phase.[16]

## IV-SUMMARY

We have outlined the temperature dependent lattice dynamics of $BiFeO_3$ from 4 K to melting by combining far infrared emissivity and reflectivity. We found that the lowest frequency unstable phonon ceases softening at about ~400 K-500 K. While it obeys below $T_N$ ~ 640 K a power law with β=0.25, as for a tri-critical point, a classical ferroelectric mode driven phase transition is frustrated by the products of $BiFeO_3$ metastability at high temperatures. We identify $T_N$ ~ 640 K as a crossover temperature above which a quasi-displacive picture driven by phonon softening is no longer valid. Above ~700 K not only thermal fluctuations but also chemical disorder is taking place partially reducing long-range order and reducing phonon life times. We have used X-ray absorption near edge structure (XANES) to find that on heating $BiFeO_3$ low temperature $Fe^{3+}$ turns into $Fe^{2+}$ while the Bi edge downshift suggests at high temperature very complex bond configuration with possible partial reduction in line with recent thermodynamic studies.[50] Overall, our local structure measurements reveal high-temperature defect induced irreversible lattice changes in an environment lacking of long-range coherence. By contrast, far infrared strong overlapping, merging, and internal mode band smearing, suggests mode tripling degeneracy in the paraelectric phase. At these wavelengths, above Tc ~1090 K, spectrum profiles may be fitted using only three oscillators thus matching the $3F_{1u}$ species of the cubic space group $Pm\bar{3}m$ (Z=1) proposed by Kamba et al for $BiFeO_3$.[20] We did not find an insulator to metal transition before melting.

## ACKNOWLEDGMENTS

N.E.M. wishes to acknowledge the friendship and hospitality of colleagues and staff at the CNRS-C.E.M.H.T.I. laboratory in Orléans, France, where all infrared measurements have been done. N.E.M. is also pleased acknowledging a sustenance grant by the Université d'Orléans, and partial founding from the Argentinean National Research Council (CONICET)-Project No. PIP 5152/06. L. D. C thanks the Basque Government for post-doctoral financial support at the CNRS-C.E.M.H.T.I. laboratory. Gilberto F. L. Fabbris and G. M A. would like to thank the Brazilian Agencies – Fundação de Amparo à Pesquisa do Estado de São Paulo (FAPESP), Conselho Nacional de Desenvolvimento Científico e Tecnológico (CNPq) and Laboratório Nacional de Luz Síncrotron (LNLS) for financial support. J. A. A. and M. J. M-L. acknowledge financial assistance from Spain Ministry of Science and Innovation (Ministerio de Ciencia e Innovación) under Project Nº MAT2007-60536.

**Permanent address:**

**§** Departamentos de Física de la Materia Condensada y Física Aplicada II, Universidad del País Vasco / Euskal Herriko Unibertsitatea, Bilbao, Apartado 644, E-48080, Spain.

26. H. M. Tütüncü and G. P. Srivastava, J. of Applied Physics, **103**, 083712 (2008).

27. G. Shirane, G., P. M. Gehring, Proceedings of the 104th Annual Meeting of the American Ceramics Society, Ceramic Transactions **136**, 17 (2003). Ibidem, J. Phys. Soc. Jpn. **84**, 5216 (2000). *April 28 May 1, 2002 in Missouri,*

28. J. F. Scott, R. Palai, A. Kumar, M. K. Singh, N. M. Murari, N. K. Karan, and R. S. Katiyar, J. A. Ceram. Soc. **91**, 1762 (2008).

29. I. Sosnowska, W. Schaefer, W. Kockelmann, K. H. Andersen, I. O. Troyanchuk, Appl. Phys. A **74**, S1040 (2002).

30. H. C. N. Tolentino, A. Y. Ramos, M. C. M. Alves, R. A. Barrea, E. Tamura, J. C. Cezar, and N. Watanabe, J. Synchrotron Rad. **8**, 1040 (2001).

31. M. Newville, J. Synchrotron Rad. **8**, 322 (2001) (http://sourceforge.net/projects/ifeffit/)

32. J. F. Brun, Ph. D. thesis, Université d'Orléans, (2004).

33. O. Rozenbaum, D. De Sousa Meneses, Y. Auger, S. Chermanne, and P. Echegut, Rev**.** Sci**.** Instrum. **70**, 4020 (1999).

34. T. Kurosawa, J. Phys. Soc. Jpn **16**, 1298 (1961).

35. Focus Software Web Site: http://crmht.cnrs-orleans.fr/pot/software/focus.html.
29

# TABLE I

Dielectric Simulation Fitting Parameters for BiFeO$_3$

| T (K) | $\varepsilon_\infty$ | $\Omega_{TO}$(cm$^{-1}$) | $\Omega_{LO}$(cm$^{-1}$) | $\gamma_{TO}$(cm$^{-1}$) | $\gamma_{LO}$(cm$^{-1}$) | $S_j$ |
|---|---|---|---|---|---|---|
| 4 K | 3.86 | 73.9 | 79.6 | 9.9 | 0.11 | 10.46 |
| | | 101.9 | 121.9 | 79.7 | 105.6 | 19.42 |
| | | 127.0 | 142.7 | 21.2 | 18.4 | 2.79 |
| | | 146.3 | 175.0 | 14.6 | 5.9 | 1.03 |
| | | 200.8 | 210.1 | 29.1 | 34.3 | 1.08 |
| | | 223.0 | 227.1 | 12.4 | 12.0 | 0.70 |
| | | 232.7 | 248.1 | 18.4 | 28.1 | 0.88 |
| | | 257.4 | 267.4 | 12.0 | 62.9 | 0.40 |
| | | 285.7 | 288.5 | 23.0 | 28.1 | 0.29 |
| | | 305.5 | 342.1 | 56.2 | 26.6 | 2.02 |
| | | 349.0 | 366.2 | 30.8 | 7.7 | 0.34 |
| | | 368.5 | 432.6 | 13.4 | 21.7 | 0.10 |
| | | 441.4 | 475.4 | 14.4 | 43.0 | 0.06 |
| | | 514.6 | 524.9 | 21.9 | 60.5 | 0.07 |
| | | 559.3 | 599.8 | 51.2 | 34.2 | 0.17 |
| 300 K | 3.30 | 69.9 | 76.1 | 11.0 | 7.7 | 8.71 |
| | | 103.5 | 128.6 | 396.8 | 593.8 | 17.27 |
| | | 129.4 | 170.7 | 56.5 | 13.2 | 0.29 |
| | | 231.2 | 235.3 | 54.0 | 49.4 | 0.65 |
| | | 254.2 | 254.9 | 326.9 | 589.6 | 0.53 |
| | | 257.5 | 269.7 | 53.0 | 127.1 | 1.70 |
| | | 275.1 | 292.5 | 369.2 | 581.1 | 0.70 |
| | | 308.6 | 348.5 | 376.2 | 54.1 | 1.32 |
| | | 349.4 | 368.9 | 52.0 | 15.6 | 0.03 |
| | | 370.6 | 430.9 | 18.4 | 52.7 | 0.04 |
| | | 440.6 | 466.7 | 36.1 | 54.4 | 0.04 |
| | | 518.0 | 524.1 | 38.0 | 58.2 | 0.05 |
| | | 560.0 | 603-4 | 80.9 | 47.1 | 0.19 |
| 523 K | 3.04 | 64.5 | 70.3 | 12.2 | 13.2 | 4.78 |
| | | 121.8 | 167.2 | 70.0 | 25.7 | 10.68 |
| | | 241.7 | 281.9 | 240.2 | 73.0 | 3.56 |
| | | 266.9 | 264.3 | 71.2 | 523.5 | 0.22 |
| | | 301.2 | 307.8 | 357.3 | 590.2 | 1.02 |
| | | 310.2 | 371.0 | 345.0 | 87.0 | 0.43 |
| | | 372.6 | 411.9 | 153.7 | 845.1 | 0,.03 |
| | | 438.9 | 466.9 | 528.2 | 85.8 | 0.14 |
| | | 530.3 | 603.1 | 106.2 | 78.3 | 0.33 |
| 850 K | 5.93 | 63.8<br>134.3<br>251.0<br>264.9<br>305.2<br>312.1<br>387.1<br>487.8<br>526.4 | 66.0<br>163.3<br>283.1<br>260.1<br>311.3<br>373.9<br>447.0<br>483.8<br>600.3 | 13.2<br>105.6<br>121.9<br>73.0<br>118.3<br>166.9<br>127.8<br>413.0<br>79.6 | 11.7<br>31.7<br>74.0<br>160.7<br>145.1<br>99.7<br>232.0<br>136.<br>113.2 | 2.3<br>11.1<br>4.86<br>1.83<br>3.08<br>0.40<br>0.48<br>0.12<br>0.62 |
| 1127 K | 4.78 | 93.0<br>207.5<br>518.3 | 153.2<br>436.8<br>599.7 | 237.5<br>101.9<br>220.3 | 69.5<br>236.8<br>152.7 | 58.5<br>13.1<br>0.52 |
| 1190 K | 4.46 | 101.3<br>229.0<br>536.2 | 163.8<br>372.5<br>622.4 | 248.7<br>134.6<br>288.8 | 156.0<br>334.9<br>143.4 | 29.84<br>6.36<br>0.92 |

# FIGURE CAPTIONS

**Figure 1. (*color online*)** X-ray (CuKα) diffraction pattern Rietveld refinement for BiFeO$_3$ at 300 K (crosses: data, full line: fit, lower full line trace: difference).

**Figure 2. (*color online*)** Temperature dependent reflectivity and emissivity of BiFeO$_3$ in the ferroelectric phase (full lines) and in the paraelectric phase (dotted lines). (a) Low temperature regime; (b) high temperature regime, the result from the melted pellet at 1279 K is also shown as a dotted line; (c) reflectivity fits at 4 K and 300 K, dots: experimental, full line: fit; (d) reflectivity and emissivity fits at 523 K and 850 K respectively, dots: experimental, full line: fit.; (e) emissivity fits at 1127 K and 1190 K in the BiFeO$_3$ paraelectric phase, dots: experimental, full line: fit. The three mode fit infrared frequencies at 1127 K are also added and outlined.

**Figure 3. (*color online*)** Temperature dependence reflectivity and emissivity of BiFeO$_3$ lowest frequency unstable phonons in the ferroelectric phase (full lines) and in the paraelectric phase (dotted lines). (a) Low temperature regime; (b) high temperature regime, the result from the melted pellet at 1279 K is also shown as a dotted line; (c) high temperature optical conductivity in the 20 cm$^{-1}$ to 180 cm$^{-1}$ frequency range; (d) temperature dependent soft phonon peak positions (dots) and power law fit (β=0.25) in the antiferromagnetic phase (full line).

**Figure 4. (*color online*)** Temperature dependent BiFeO$_3$ Bismuth L$_{III}$-edge XANES spectra. on warming.

**Figure 5. (*color online*)** Temperature dependent BiFeO$_3$ Iron K-edge XANES spectra.

**Figure 6. (*color online*)** XANES spectra comparison at room temperature of BiFeO$_3$ Iron K-edge cooled from 1123 K (full line), FeO (dashed line), Fe$_3$O$_4$ (circle), and Fe$_2$O$_3$ (square). Inset: EXAFS pseudo radial distribution function of BiFeO$_3$ (full line) and FeO (dashed line).

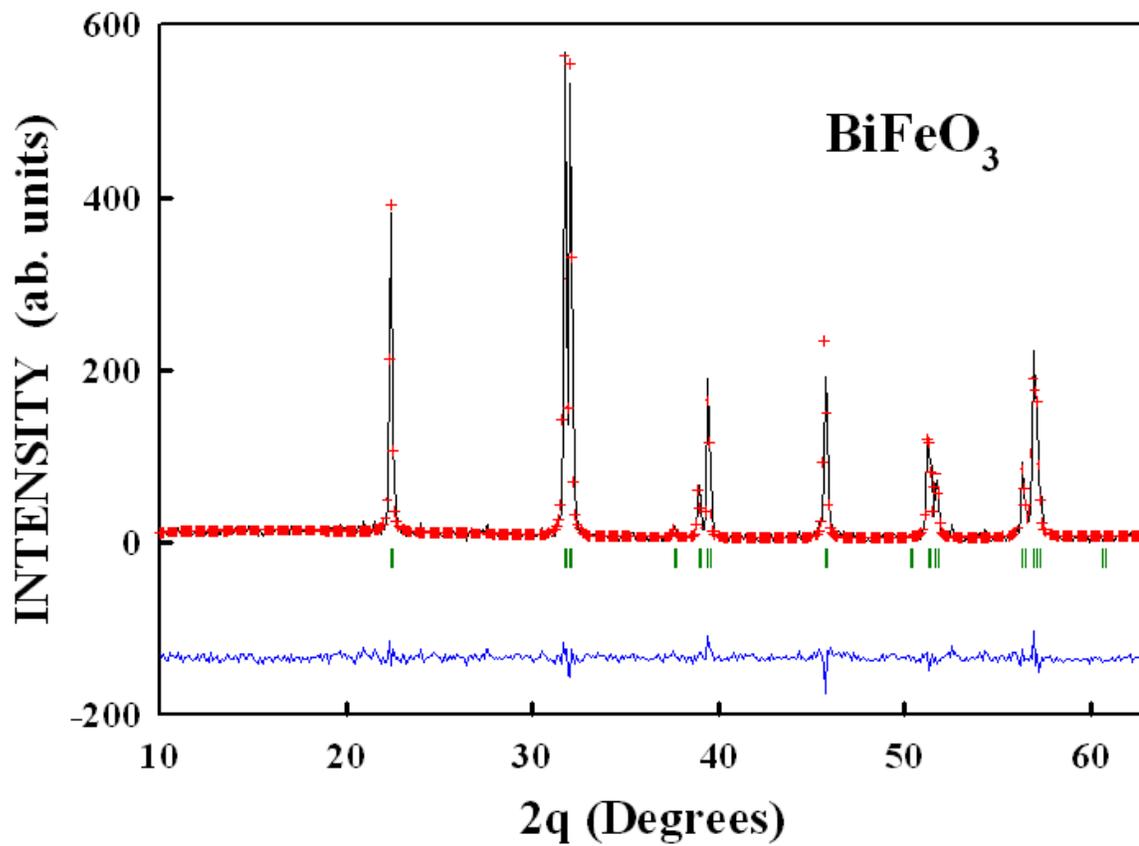

Figure 1
Massa et al

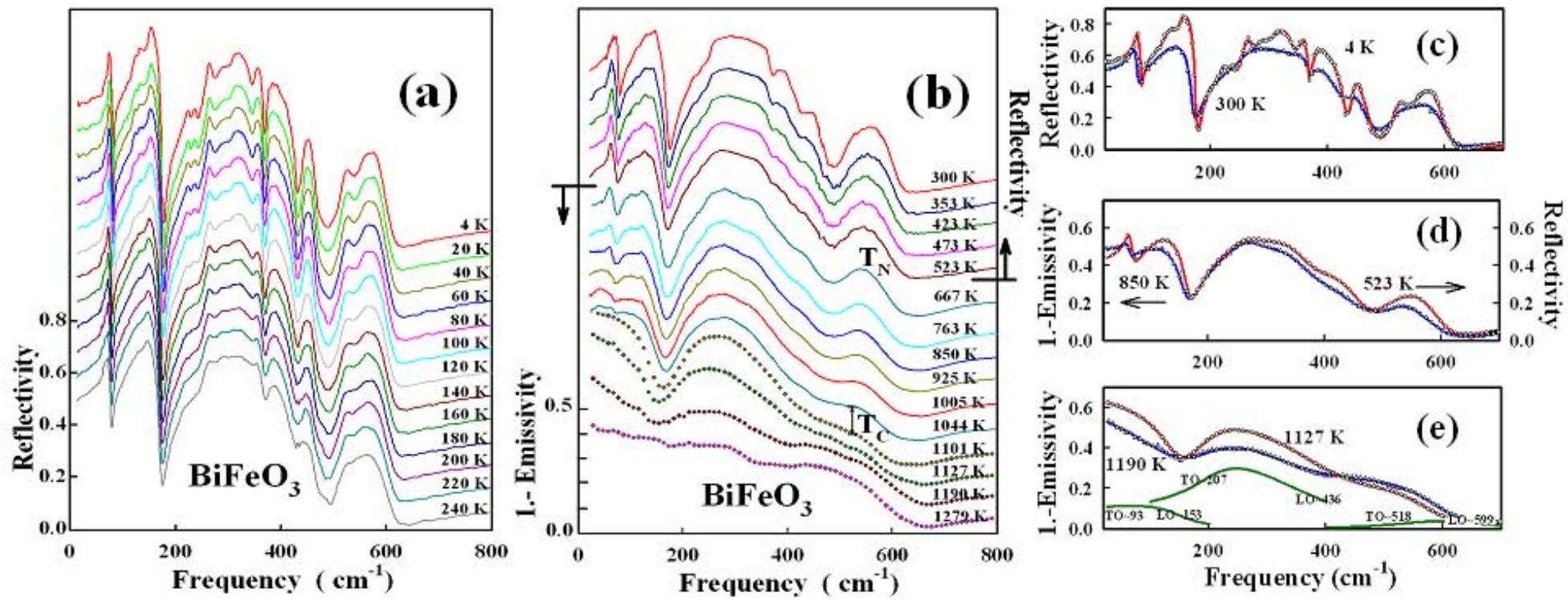

Figure 2
Massa et al

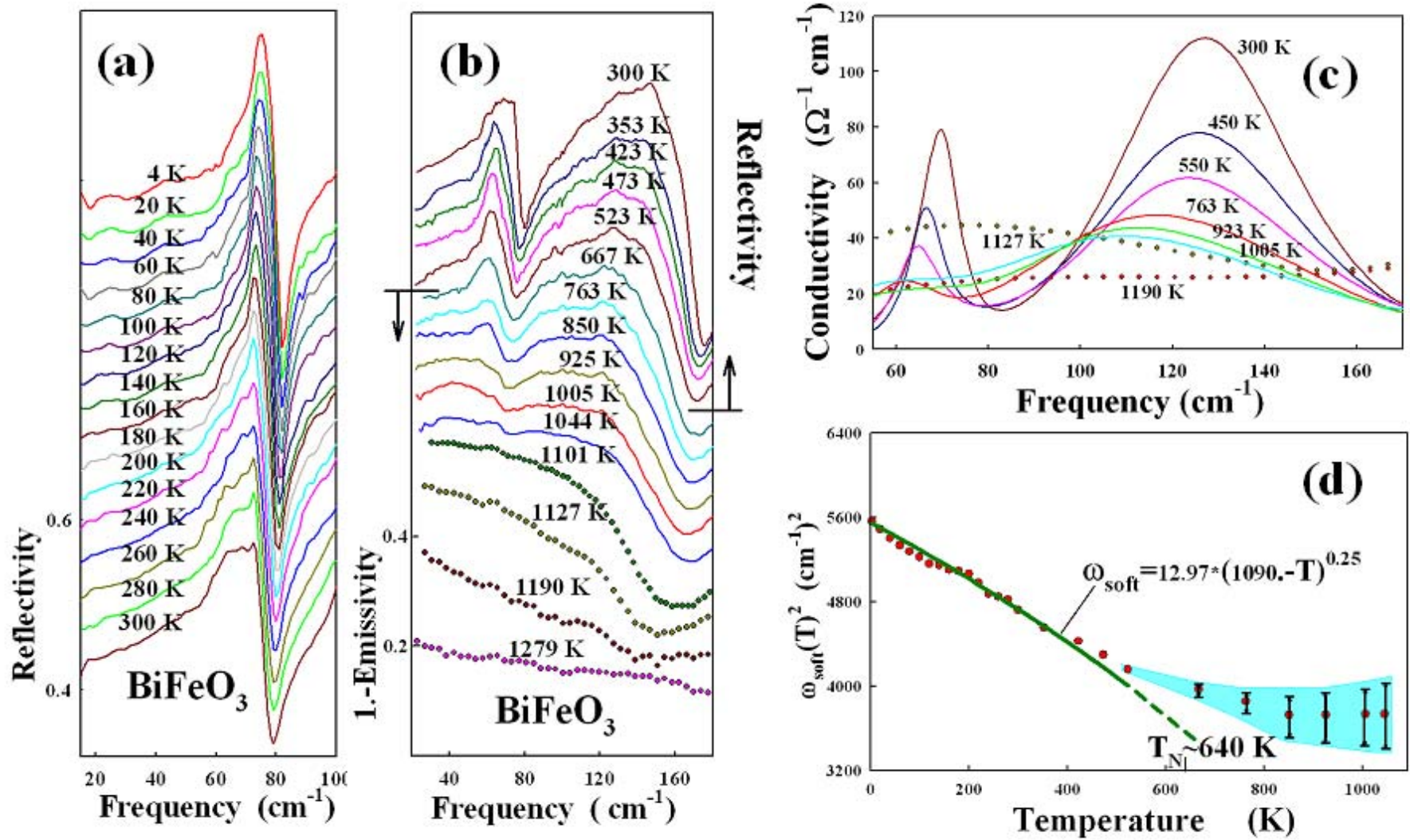

Figure 3
Massa et al

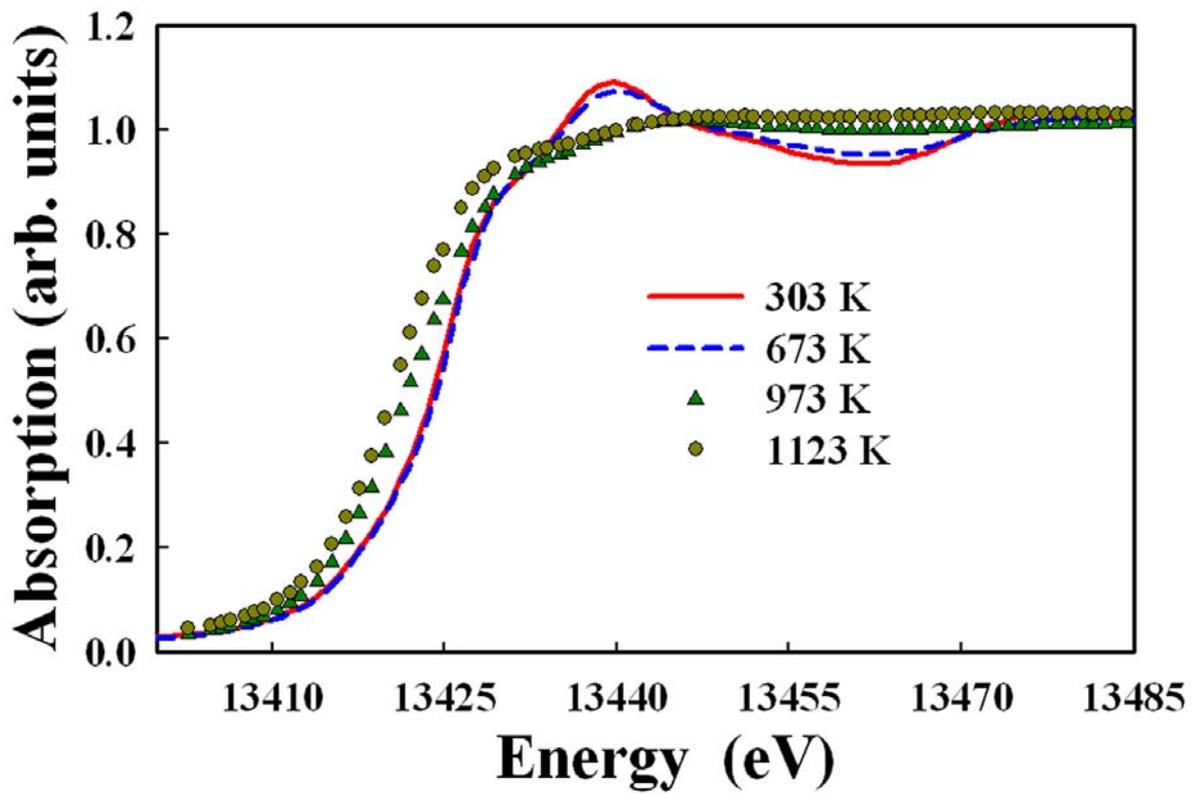

**Figure 4
Massa et al**

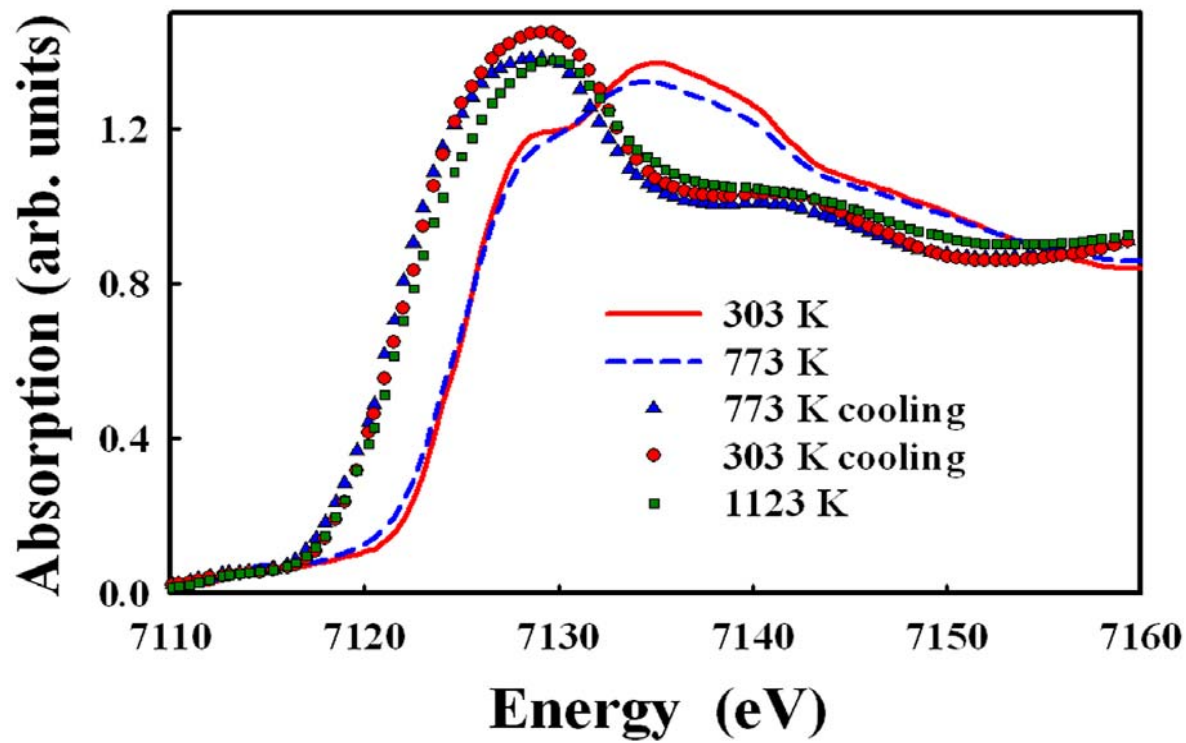

**Figure 5
Massa et al**

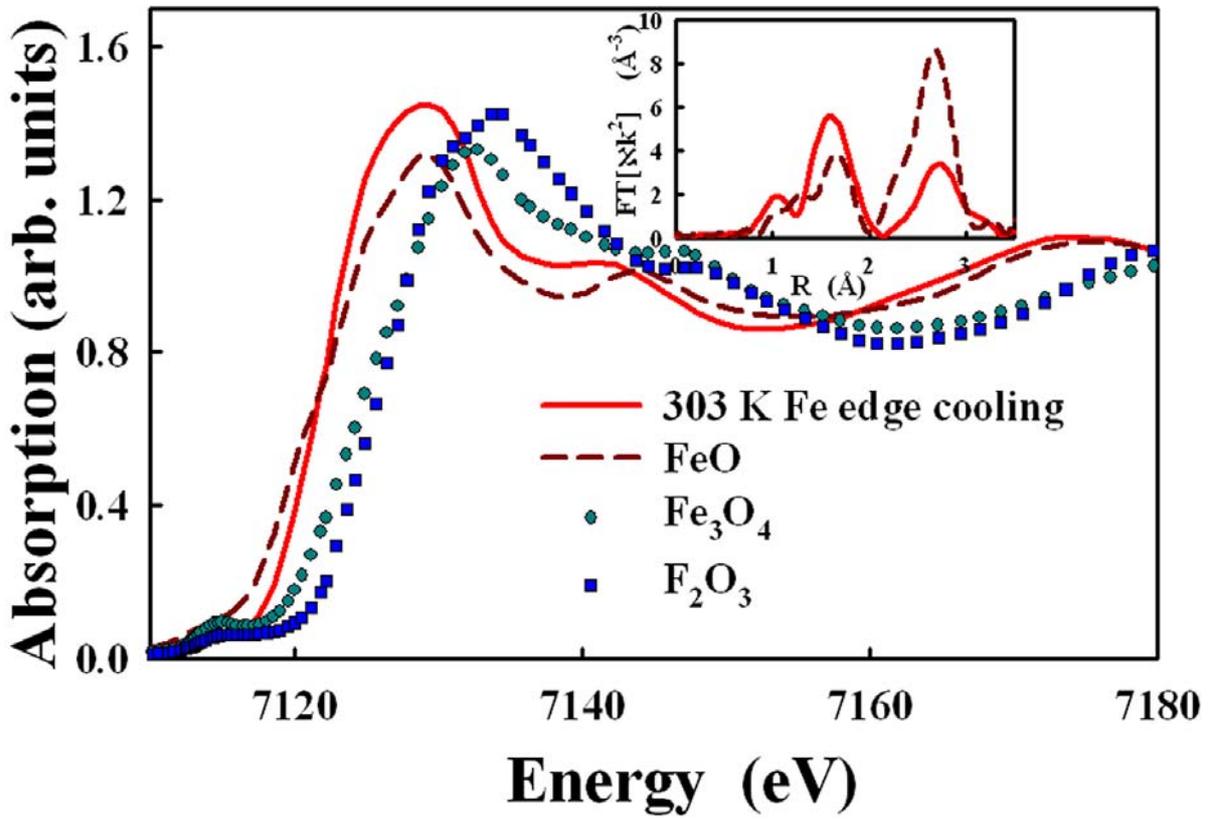

**Figure 6
Massa et al**